\documentclass[aps,pra,showpacs,twocolumn]{revtex4}
\usepackage{amssymb}
\usepackage{epsfig}
\usepackage{graphicx}
\usepackage{amsmath}
\usepackage{array,color}
\usepackage{dcolumn}
\usepackage{bm}

\begin{document}
\title{Kondo metal of fermionic atoms in anisotropic triangular optical lattice}
\author{Baoan$^1$}
\author{Xiaozhong Zhang$^1$}
\author{Wu-Ming Liu$^2$}
\affiliation{
$^1$Laboratory of Advanced Materials, Department of Materials Science and Engineering, Tsinghua University, Beijing 100084, China\\
$^2$Beijing National Laboratory for Condensed Matter Physics, Institute of Physics, Chinese Academy of Sciences, Beijing 100190, China}
\date{\today}
\begin{abstract}
Quantum phase transition of fermionic atoms in anisotropic triangular optical lattice was investigated by dynamical cluster approximation combining with continuous time quantum Monte Carlo algorithm. The temperature-interaction phase diagram for different hoping terms and the competition between the anisotropic parameter and interaction is presented. Our results show that the system undergoes Mott transition from Fermi liquid to Mott insulator while the repulsive interaction reach the critical value. The Kondo metal characterized by Kondo peak in the density of states is found in this systems and the pseudogap are suppressed at low temperature due to the Kondo effect. We also propose a feasible experiment protocol to observe these phenomenon in the anisotropic triangular optical lattice with the cold atoms, in which the hoping terms can be varied by the lattice depth and the atomic interaction can be adjusted via Feshbach resonance.
\end{abstract}
\pacs{03.75.Ss, 37.10.Jk, 32.80.Hd, 67.85.Lm}
\maketitle
\section{Introduction}
As an ideal toolbox to investigate the quantum many-body phenomenon encountered in condensed matter physics, cold atoms in optical lattices with perfect structure and highly controllable parameters have attracted great attentions recently \cite{Petsas,Greiner1,Jaksch,Immanuel}. Many cold atoms in optical lattice based quantum simulation work have been done with fruitful results, such as the observations of the Mott insulator and superfluidity of strongly interacting fermions in optical lattice and the quantum phase transition between these novel stases, the quantum simulations of the antiferromagnetic spin chains in optical lattice, the frustrated Classical Magnetism in triangular optical lattice, Bose-Einstein condensates in optical lattices so forth \cite{Jordans,Gemelke,Helmes,Hofstetter,Chin,Cheng,Greiner,Simon,Struck,Konotop}. Optical lattice with different structure and dimension can be set up by overlapping several beams of laser with the same frequency and different propagating directions \cite{Duan,Duan05,Wu,Yao,Santos,Schneider}. In terms of the setting-up of the cold atoms in the triangular optical lattice which is geometrically equilateral but anisotropic in the hoping terms, three pairs of laser with the same frequency and different intensity in the same plane are used to create periodic potentials, as shown in Fig. 1. The fermionic $^{40}K$ atoms in the two hyperfine manifold are loaded into the lattice potential by combining different cooling methods including laser cooling and evaporative cooling. The interactions between fermions can be tuned by Feshbach resonance \cite{Cheng} and the hoping between the nearest neighbor sites are closely related to the confining potential depth in different directions.

To our knowledge, the novel states including the spin liquid state and the spin ice which found in the frustrated systems play pivotal role in the understanding of the mechanism of the high temperature superconductivity and magnetic monopoles. Hence, many experimental and theoretical studies have focused on geometrically frustrated systems \cite{Shimizu,Harris,Kurosaki,Castelnovo,Ohashi,Yoshioka}. However, there is few work on the anisotropic triangular systems which is equilateral in the lattice constant but anisotropic in the hoping terms in different directions.

In this paper, we investigate the fermionic cold atoms in a anisotropic triangular optical lattice by combining the dynamical cluster approximation (DCA) \cite{Maier} with the continuous time quantum monte carlo method. Even though the dynamical mean-field theory (DMFT) is proved to be exact in the infinite-dimensional limit and a good approximation even for three dimensions \cite{Metzner}, the DCA has some advantages over than the DMFT to study frustrated systems, because the DCA can reasonably take into account the nonlocal correlations of the frustrated systems which is overlooked in the DMFT. In the dynamical cluster approximation method, the lattice problem is mapped into a self-consistently embedded finite-sized cluster and the irreducible quantities of the embedded cluster are used as an approximation for the corresponding lattice quantities. This paper is organized as follows. In Sec. II, the artificial anisotropic triangular lattice with femionic atoms, the algorithm and method which are used to describe such artificial systems are discussed in detail. In Sec. III, we present the  double occupancy and the density of state of the system. The Fermi surface evolution, the temperature-interaction (T-U) phase diagram  and the competition between the anisotropic parameters $\lambda$ and interaction are also shown in Sec. III. In Sec. V,  We design a feasible experiments for the observation of the quantum phase transitions of the cold fermionic atoms in the anisotropic triangular optical lattice that is geometrically equilateral but anisotropic in hoping terms. A brief discussion and conclusion is presented in the last section.
\section{ANISOTROPIC TRIANGULAR OPTICAL LATTICE}
\begin{figure}[!t]
\centering
\includegraphics[width=9.cm]{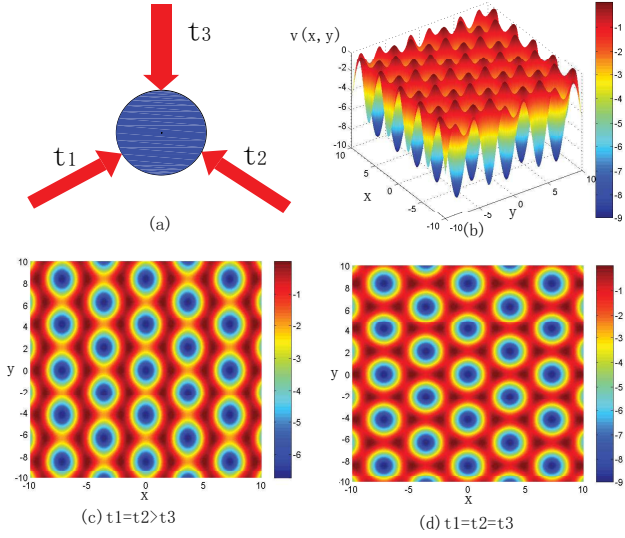}\hspace{0.7cm}
\caption{\label{fig:epsart}(Color online)
(a): Sketch map of experimental setup to form anisotropic triangular optical lattice, the degree of thickness of directed arrow represent the intensity of the laser beams and each arrow depicts the propagating laser beam and its direction, the circle in center of the figure represent the fermions.
(b): cubic sketch of potentials $V(x,y)$ in space for the case of isotropic triangular lattice. (c): plane sketch of potentials distribution of anisotropic triangular while $t_3$$<$1 and $t_1$=$t_2$=1. (d): plane sketch of potentials distribution of anisotropic triangular while $t_1$=$t_2$=$t_3$=1.}
\end{figure}
Artificial anisotropic triangular optical lattice which is geometrically equilateral in lattice constant but anisotropic in the hoping terms of different directions can be created through the interference of three pairs of laser with the same wavelength $\lambda=1064$ nm and different intensity in the same plane. The included angel between each two standing wave laser should be $2\pi/3$. The fermionic $^{40}K$ atoms in the two hyperfine manifold are loaded into the triangular optical lattice in the same percentage through evaporative cooling at a magnetic bias field of 203.06 G \cite{Jordans}. The hoping terms between nearest neighbor sites in different directions are closely related to the combining potentials depth $V_0$ which adjusted by the lasers' intensity. The anisotropy in this equilateral triangular lattice can be realized through the manipulation of the hoping term t1, t2 and t3 in the different direction, as shown in Fig. 2(a). Both the non-interacting regime with a scattering length of a = (0 $\pm$ 10)$a_0$ and the repulsive on-site interactions between fermionic atoms can be tuned by Feshbach resonances, where $a_0$ is the Bohr radius. The confining potential in space is
\begin{eqnarray}\label{eq:eps}
V(x,y)&=&V_0\{[t_1\cos(\alpha)+t_2\cos(\beta)+t_3\cos(\gamma)]^2\nonumber\\
      &+&[t_1\sin(\alpha)+t_2\sin(\beta)+t_3\sin(\gamma)]^2\},
\end{eqnarray}
where $V_0$ is potential depth in the x-y plane and is given in unit of recoil energy is $\emph{E}_r$=$h^2$/(2m$\lambda^2$), here h is Planck's constant, m is the atomic mass and $\lambda$ is the wavelength of the laser beam. t1, t2 and t3 are the hoping term in different direction respectively. $k_x$ and $k_y$ are the two components of wave vector $k=2\pi/\lambda$ along x and y directions. $\alpha$=$(\tfrac{-\sqrt{3}k_x}{2}+\tfrac{k_y}{2})$, $\beta$=$\tfrac{-\sqrt{3}k_x}{2}-\tfrac{k_y}{2}$, $\gamma$=$k_y$.

\begin{figure}[!t]
\centering
\includegraphics[width=6.0cm]{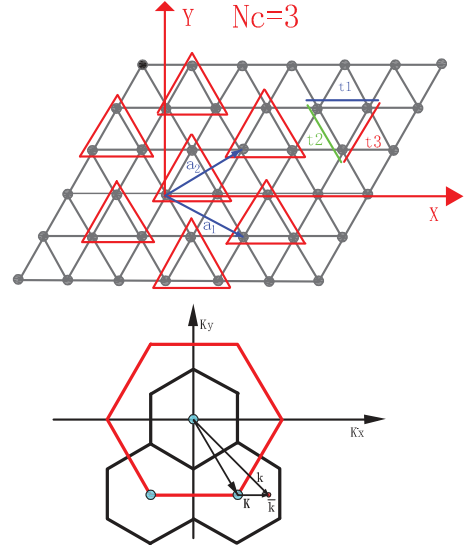}\hspace{0.5cm}
\caption{\label{fig:epsart}(Color online)
(a): Sketch of anisotropic triangular lattice. t1, t2 and t3 represents the hoping term of different directions respectively
(b): Coarse graining procedure in the first Brillouin Zone when $N_c=3$. The red part shows the first Brillouin Zone.
The 3 black hexagonals show the region of divided Brillouin Zone for DCA calculation.}
\end{figure}

The fermionic cold atoms with repulsive on-site interaction in anisotropic triangular optical lattice can be described by Hubbard model and the Hamilitonian is
\begin{equation}\label{eq:eps}
H=-\sum_{<i,j>\sigma}t_{ij}c_{i\sigma}^{+}c_{j\sigma}+U\sum_{i}n_{i\uparrow}n_{i\downarrow},
\end{equation}
where $c_{i\sigma}^{+}$ and $c_{i\sigma}$ represent the creation and the annihilation operator of fermios on lattice site respectively, $n_{i\sigma}=c_{i\sigma}^{+}c_{i\sigma}$ is density operator of fermions. The general formula for the hoping terms is  $t=(4/{\sqrt{\pi}})E_r(V_0/E_r)^{3/4}exp(-2(V_0/E_r)^{1/2})$. The different hoping terms in different directions can be achieved by adjusting the lattice depth $V_0$ in different directions. $U=\sqrt{8/\pi}ka_sE_r(V_0/E_r)^{3/4}$ is the on-site repulsive interaction which can be tuned by Feshbach resonance.

In dynamical cluster approximation the lattice problem is mapped into a self-consistently embedded finite-sized cluster and the reciprocal space of the lattice containing $N$ points is divided into finite cells. The coarse-graining Green's function $\overline{G}$ is achieved by averaging Green's function $G$ within each cell. The coarse graining procedure of anisotropic triangular lattice in the DCA is illustrated as follows. As Fig. 2 shows that the lattice problem of anisotropic triangular lattice is mapped onto three sites cluster, the Brillouin zone is divided into 3 cells and each cell is represented by a cluster momentum $\mathbf{K}$.
The coarse-grained Green's function is
\begin{eqnarray}
\overline{G}(\textbf{K},i\omega_n)=\frac{Nc}{N}\sum_{\widetilde{\textbf{k}}}\frac{1}{i\omega_n-\varepsilon_{\textbf{K}+
\widetilde{\textbf{k}}}-\overline{\Sigma_\sigma}(\textbf{K},i\omega_n)},
\end{eqnarray}
where summation over $\widetilde{\textbf{k}}$ is taken within the coarse-graining cell, the $\omega_n$ is the Matsubara frequency.

After mapping the Hubbard model onto finite size cluster, we use the  continuous time quantum monte carlo method (CTQMC) as impurity solver to deal with the
cluster problem. The CTQMC does not use the Trotter decomposition, so it is more exact than the general QMC. We did $10^7$ sweeps in our CTQMC step. The process from
$\overline{G}(\textbf{K},i\omega_n)=\frac{Nc}{N}\sum_{\widetilde{\textbf{k}}}1/(i\omega_n-\varepsilon_{\textbf{K}+
\widetilde{\textbf{k}}}-\overline{\Sigma_\sigma}(\textbf{K},i\omega_n))$ to $\Sigma_c(\textbf{K},i\omega_n)=\overline{\mathcal{G}}(\textbf{K},i\omega_n)^{-1}-\overline{G}(\textbf{K},i\omega_n)^{-1}$
have been repeated until $\Sigma_c(\textbf{K},i\omega_n)$ converges to required accuracy, as shown in Fig. 3.

\begin{figure}[!t]
\centering
\includegraphics[width=8cm]{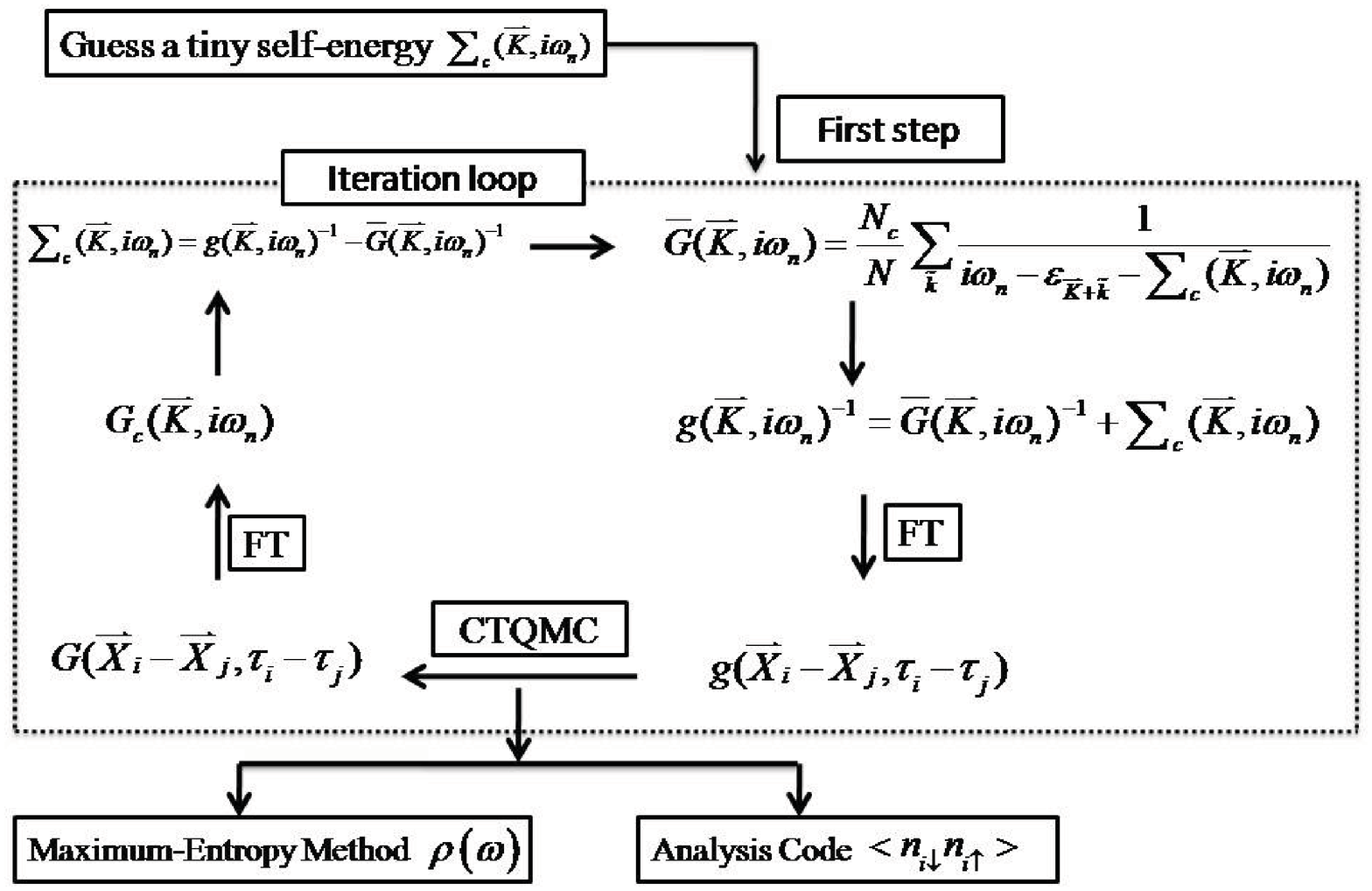}\hspace{0.5cm}
\caption{\label{fig:epsart} The flow diagram of continuous time quantum monte carlo iteration loop. FT represents fourier transformation.}
\end{figure}

\section{PHASE DIAGRAM}
We calculate the double occupancy and the density of state which are important to the study of the quantum phase transition in optical lattice systems. We also show the evolution of the Fermi surface's evolution at fixed temperature and repulsive interactions and the phase diagram of the system.

The double occupancy $D_{occ}=\partial{F}/\partial{U}=\frac{1}{4}\sum_{i}\langle{n_{i\uparrow}n_{i\downarrow}}\rangle$ as a function of interaction $U$ for various temperature and different hoping terms has been investigated, where F is the free energy. Here we define anisotropic parameter $\lambda$ = $t_3$/$t_i$ (i = 1,2), which describes the degree of anisotropy of triangular optical lattice. The double occupancy (Docc) increases with the increase of the $\lambda$ = $t_3$/$t_i$ at fixed interaction and temperature. With the increase of interaction the effect of temperature on Docc is weakened due to the suppressing of the itinerancy of the atoms and suppressed while $U/t_i$ is larger than 10.0 and 12.0 for $\lambda$ $<$ 1 and $\lambda$ $>$ 1 respectively. As Fig. 4 shows that the Docc of the systems don't overlap in the fermi liquid phase.

\begin{figure}[!t]
\centering
\includegraphics[width=8.0cm]{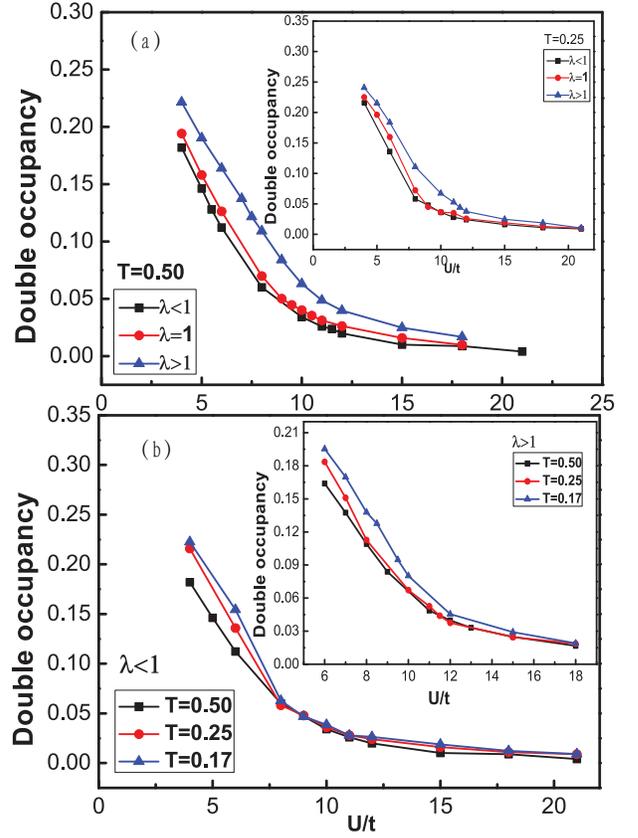}\hspace{0.5cm}
\caption{\label{fig:epsart}(Color online) The double occupancy $D_{occ}$ as a function of the interaction U for different temperature and $\lambda$ has been shown, where $\lambda$ =$t_3/t_{1,2}$ and $t_1$=$t_2$. (a): The double occupancy against interaction for fixed temperature at different anisotropic parameter $\lambda$. (b):  The double occupancy against interaction for fixed anisotropic parameter $\lambda$ at different temperature.}
\end{figure}

The density of state $\rho(\omega)$ (DOS) are calculated by maximum entropy method \cite{Gubernatis}. At T=0.5, the formation process of a pseudogap for $t_3$=0.8 and $t_3$=1.5 is different. For $t_3$=0.8, a pseudogap directly formed by splitting Fermi-liquid-like peak at $U/t_i$=7.0 without the appearance of the Kondo peak, while for $t_3$=1.5, a pseudogap appears following after the appearance of the Kondo peak at $U/t_i$=10.0. Still at T=0.5, the system transferred to insulator state while $U_c/t_i=10.0$ and $U_c/t_i$=12.0 for $t_3$=0.8 and $t_3$=1.5 respectively. At T=0.25, Kondo resonance peak which characterized by a sharp quasi-particle peak with two shoulders appears at $U_c/t_i$=6.5 and $U_c/t_i$=8.5 respectively at $t_3$=0.8 and $t_3$=1.5. A gap is opened when the interaction is stronger than the critical interaction $U_c/t_i$=9.5 for $t_3$=0.8 and $U_c/t_i$=11.8 for $t_3$=1.5. The Kondo peak formed at low temperature because of the effect between the local atom and the itinerant atom instead of that formed at higher temperature through splitting the Fermi-liquid like into two parts.
\begin{figure}[t]
\centering
\includegraphics[width=8.5cm]{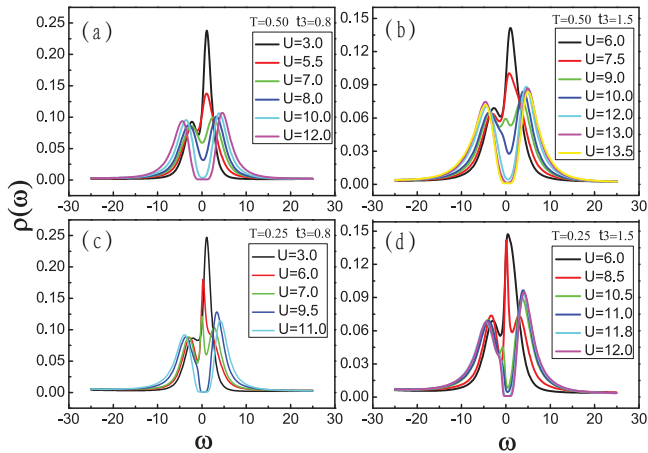}\hspace{0.5cm}
\caption{\label{fig:epsart}(Color online) The density of states as a function of frequency $\omega$ for different interaction U/$t_i$ (i=1,2). (a): The density of states against frequency $\omega$ for T=0.5 and $t_3$=0.8. (b): The density of states against frequency $\omega$ for T=0.5 and $t_3$=1.5. (c): The density of states against frequency $\omega$ for T=0.25 and $t_3$=0.8. (d): The density of states against frequency $\omega$ for T=0.25 and $t_3$=1.5.}
\end{figure}

The Fermi surface's evolution for different hoping terms $t_3$ at fixed $t_1$=$t_2$=1 is shown in Fig. 6 while T/$t_i$=0.25 and $U/t_i$=8.0 (i=1, 2). The spectral function is $A(k;\omega=0)\approx-\frac{1}{\pi}\lim_{w_n\rightarrow0}ImG(k,i\omega_n)$. A linear extrapolation of the first two Matsubara frequencies is used to estimate the self-energy to zero frequency\cite{Parcollet}. Due to the delocalization of the particles the amplitude of the spectral weight becomes larger and the ring breadth becomes narrow and distinct with the increase of the hoping terms for fixed temperature and interaction. The Fermi surface becomes a nearly flat plane when $t_3$ is lower than 1.0 at T=0.25. The transition from Fermi liquid to Mott insulator is much faster at low temperature than that at relatively high temperature.
\begin{figure}[!t]
\centering
\includegraphics[width=9cm]{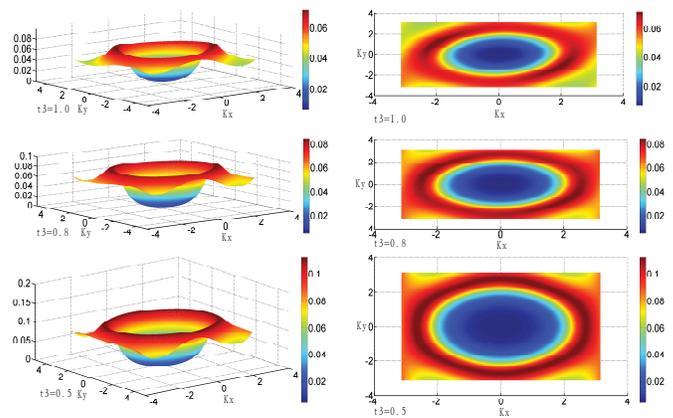}\hspace{0.5cm}
\caption{\label{fig:epsart}(Color online) The Fermi surface as a function of momentum $k$ for the different hoping terms $t_3$=0.5, 0.8, 1.0 at fixed $t_1$=$t_2$=1, $T$/$t_i$=0.25 and $U$/$t_i$=8.0 (i=1, 2).}
\end{figure}

The temperature-interaction (T-U) phase diagram and the competition between anisotropic parameters $\lambda$ and interaction of cold atoms in anisotropic triangular optical lattice are shown in Fig. 7. With the increase of the hoping terms those corresponding transition points entirely shifted to larger interaction and lower temperature. For t3=0.8, there is no Kondo peak between Fermi liquid and pseudogap until the T is lower than 0.45. With the increase of the hoping term $t_3$, at fixed $t_1$=$t_2$=1, Kondo peak not only appears but the gap between Kondo peak pseudogap becomes nearly zero. When the interaction is stronger than the critical value of interaction for Mott transition, the system translates into Mott insulator regime which confirmed by an opened gap. The pseudogap are suppressed at low temperature due to the Kondo effect. The competition between anisotropic parameter $\lambda$ and interaction at T/$t_i$=0.5 in Fig. 7 (d) shows that with the increase of the interaction,the transition points $\lambda$ for the fermi liquid to Mott insulator increase and the slope decreases, respectively.

\section{EXPERIMENTAL PROTOCOL}

We propose an experiment scheme for the observation of the quantum phase transition of fermionic cold atoms in geometrically frustrated anisotropic triangular optical lattice. First, we use three pairs of laser in the same plane at wavelength $\lambda=1064$ nm with different intensity to create optical lattice. The included angle between each two beams of laser equals $2\pi/3$ \cite{Tung}. A balanced mixture of the fermionic $^{40}K$ atoms in two hyperfine manifold states, $\mid$$m_F$$>$=$\mid$-9/2$>$ and $\mid$-7/2$>$ states, is prepared through evaporative cooling at a magnetic bias field \cite{Jordans,Ohara}. Then a spin mixture of atoms in the $\mid$-9/2$>$ and $\mid$-5/2$>$ states can be loaded into the optical lattice potentials through tuning the scattering length to the desired value by Feshbach resonance \cite{Regal,Klempt,Thilo,Loftus,Zwierlein}. We can achieve the zero interaction regime with a scattering length of a=(0$\pm$10)$a_0$, where $a_0$ is Bohr radius. The repulsive on-site interaction $U/t_i$ between fermionic atoms can be obtained by transferring the atoms in $\mid$-7/2$>$ state to the $\mid$-5/2$>$ state during the evaporation cooling.
With the increase of the interactions the system will transform from Fermi liquid ($a_s$$<$48$a_0$) to Mott insulator($a_s$$>$111$a_0$) at $T$=5.96nK for the lattice depth $V_0$=10$E_r$. $a_s$ indicates the s-wave scattering length which is used to determine the effective interaction region. In order to detect the double occupancy, further tunnelling should be prevented through increasing the potential depth $V_0$ of optical lattice rapidly and the energy of atoms on doubly occupied sites need to be shifted by approaching Feshbach resonance. One spin component of fermions on the double occupied sites is transferred to an unpopulated magnetic sublevel by using a radio-frequency pulse. The double occupancy can be deduced by the fraction of transformed atoms obtained through the absorption imagining \cite{Jordans, Strohmaier}. After completely turning off the confining potential, the atoms expand anisotropically for several milliseconds due to the anisotropy of the potential depth in different direction. One can get the Fermi surface through the absorption imagining \cite{Chin,Kohl}.
\begin{figure}[!t]
\centering
\includegraphics[width=8.5cm]{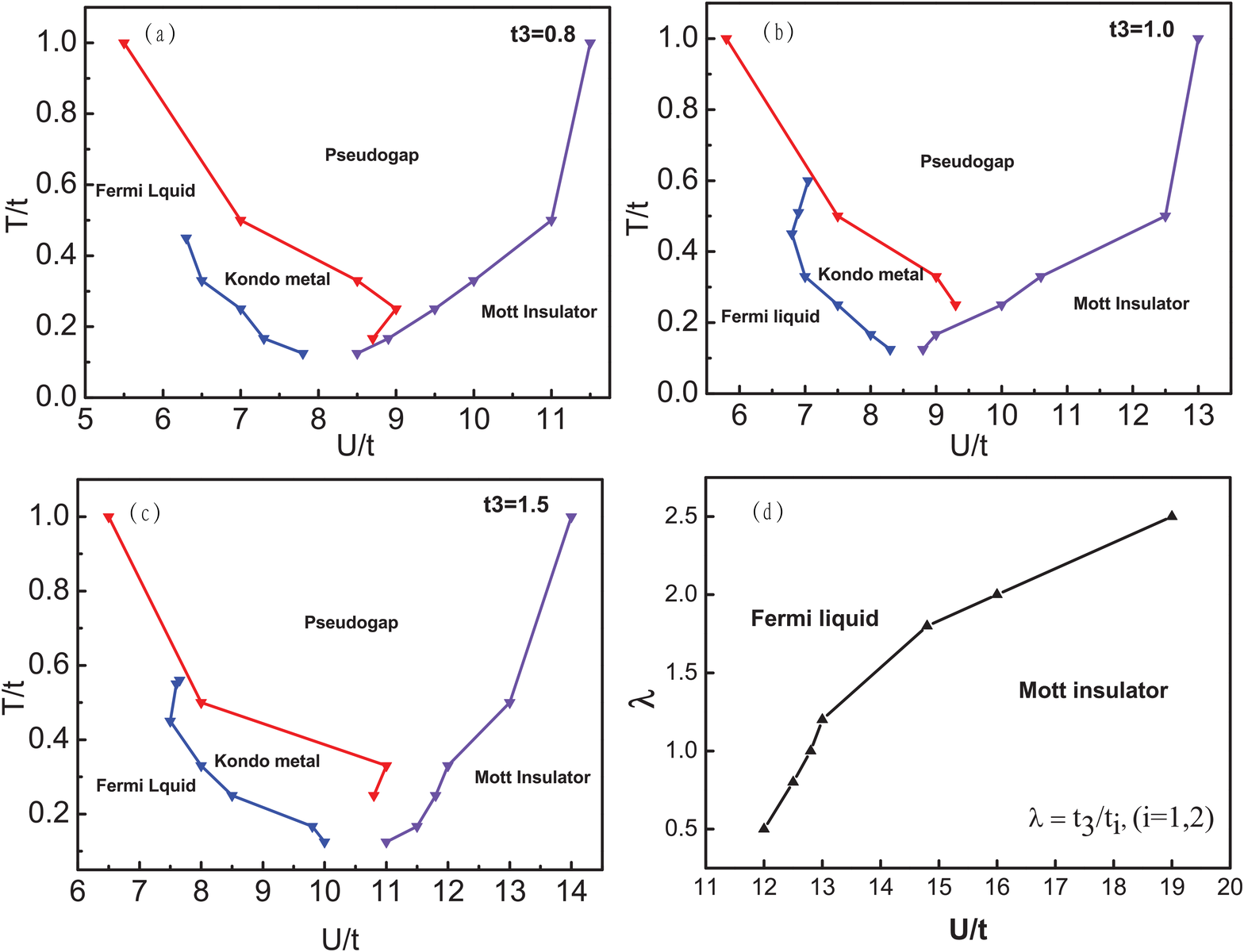}\hspace{0.5cm}
\caption{\label{fig:epsart}(Color online) (a): The temperature-interaction phase diagrams of fermionic cold atoms in anisotropic triangular optical lattice for $t_3$=0.8. (b): The temperature-interaction phase diagrams of fermionic cold atoms in anisotropic triangular optical lattice for $t_3$=1.0. (c): The temperature-interaction phase diagrams of fermionic cold atoms in anisotropic triangular optical lattice for $t_3$=1.5. (d): The competition between $\lambda$=$t_3$/$t_i$ (i=1, 2) and interaction U.}
\end{figure}
\section{SUMMARY}
We investigate the quantum phase transitions of fermionic cold atoms in geometrically frustrated anisotropic triangular optical lattice with spin half fermions by dynamical cluster approximation method combining with the continuous-time quantum monte carlo algorithm in the frame of half-filled and one band Hubbard model. The system transform from Fermi liquid into Mott insulator while the repulsive on-site interaction reaches certain value for different temperature. We also found Kondo metal phase at middle regime temperature. The Pesudogap is totally suppressed at low temperature due to the Kondo effect. It is found that with the increase of $t_3$ at fixed $t_1$=$t_2$, the transition lines shift to larger interaction and lower temperature. In the Mott insulator regime, the effect of the repulsive interaction on double occupancy becomes unapparent.

This work was supported by NKBRSFC under grants Nos. 2011CB921502, 2012CB821305, 2010CB922904, 2009CB930701, 2009CB929202, by NSFC under grants Nos. 11074141, 11174169, 10934010 and by NSFC-RGC under grants No. 11061160490.

\end{document}